\documentclass[11pt]{amsart}
\usepackage[usenames]{color}

\usepackage{latexsym}
\usepackage{amsfonts}
\usepackage{amsthm}
\usepackage{graphicx}
\usepackage{amsmath}
\usepackage{amssymb}
\newtheorem{definition}{Definition}

 \newtheorem{problem}{Problem}
 
 \theoremstyle{definition}
 \theoremstyle{remark}

 \numberwithin{equation}{section}

\begin{document}

\title[A Secret Sharing Scheme Based on Group Presentations]
{A Secret Sharing Scheme Based on Group Presentations and the Word
Problem}
\author[M. Habeeb]{Maggie Habeeb}
\address{CUNY Graduate Center, City University of New York}%
\email{MHabeeb@GC.Cuny.edu}%
\author[D. Kahrobaei]{Delaram Kahrobaei}
\address{CUNY Graduate Center, City University of New York}%
\email{DKahrobaei@GC.Cuny.edu}%
\author[V. Shpilrain]{Vladimir Shpilrain}
\address{The City College of New York and CUNY Graduate Center}
\email{shpil@groups.sci.ccny.cuny.edu}

\maketitle

\begin{abstract}
A $(t,n)$-threshold secret sharing scheme is a method to distribute
a secret among $n$ participants in such a  way that any $t$
participants can recover the secret, but no $t-1$ participants can.
In this paper, we propose two secret sharing schemes using
non-abelian groups.  One scheme is the special case where all the
participants must get together to recover the secret.  The other one
is a $(t,n)$-threshold scheme that is a combination of Shamir's
scheme and the group-theoretic scheme proposed in this paper.
\end{abstract}


\section{Introduction}

Suppose one would like to distribute a secret  among $n$
participants in such a way that any $t$ of the participants can
recover the secret, but no  group of $t-1$ participants can.  A
method that allows one to do this is called a $(t,n)$-threshold
scheme.  Shamir \cite{Shamir} back in 1979 offered a
$(t,n)$-threshold scheme  that utilizes polynomial interpolation. In
Shamir's scheme the secret is an element $x\in \mathbb{Z}_{p}$.  In
order to distribute the secret, the dealer begins by choosing a
polynomial $f$ of degree $t-1$ such that $f(0)=x$.  Then he sends
the value $x_{i}=f(i)$ secretly to participant $P_{i}$. In order for
the participants to recover the secret, they use polynomial
interpolation to recover $f$ and hence the secret $f(0)$.  In this
setting, no $t-1$ participants can gain any information about the
secret while any $t$ of them can (see \cite{Shamir} or
\cite{drstin}).

The field of non-commutative group-based cryptography has produced
many new cryptographic protocols over the last decade or so.  We
refer an  interested reader to \cite{MSU} or \cite{FHKR} for a
survey of developments in this area. Recently,  Panagopoulos
\cite{DP} suggested a $(t,n)$-threshold scheme using group
presentations and the word problem.
His scheme is two-stage: at the first stage,  long-term private
information (defining relations of groups) is distributed to all
participants over secure channels. At the second stage, shares of
the actual secret are distributed to participants over open
channels. Thus, an advantage of Panagopoulos' scheme  over Shamir's
scheme is that the actual secret need not be  distributed over
secure channels. This is obviously useful in various real-life
scenarios. On the other hand, his scheme is not quite practical
because in his scheme, it takes time exponential in $n$ to
distribute the shares of a secret to $n$ participants.

In this paper, we present two practical secret sharing schemes based
on group presentations and the word problem.   The first one is a
two-stage scheme, like the one due to Panagopoulos mentioned above,
but it is just  an $(n,n)$-threshold scheme. Our second scheme is a
$(t,n)$-threshold scheme, $0<t\leq n$. Both our schemes are designed
for the scenario where the dealer and participants initially are
able to communicate over secure channels, but afterwards they
communicate over open channels.

We first consider the special (trivial) case where $t=n$; that is,
we propose a scheme in which all participants are needed to recover
the secret.
%
%
%
Then, we  propose a hybrid scheme that combines Shamir's scheme and
the idea of the $(n,n)$-threshold scheme we proposed. This combined
scheme  has the same distributed secret as Shamir's scheme does, but
rather than sending $f(i)=x_{i}$ over secure channels we send the
integers in disguise, and then participants use group-theoretic
methods to recover the integers.  This scheme has the following
useful advantages over Shamir's scheme:

\begin{itemize}
\item The actual secret need not be  distributed over
secure channels and, furthermore, once the long-term private
information is distributed to all participants, several different
secrets can be distributed without updating the  long-term private
information.


\item While recovering the secret,  participants do not have to reveal their shares
to each  other if they do not want to.
\end{itemize}

\section{Very Brief Background on Group Theory}
\label{back}

In this section, we give a minimum of information and notation from
group theory necessary to understand our main Sections \ref{(n,n)}
and \ref{(t,n)}. Further facts from group theory that are used,
explicitly or implicitly, in this paper are collected in Sections
\ref{Platform} and \ref{Tietze}.

A free group $F=F_m$ of rank $m$, generated by $X=\left\{x_{1},
\ldots, x_{m}\right\}$, is the set of all reduced words in the
alphabet $\left\{x_{1}^{\pm 1}, \ldots, x_{m}^{\pm 1}\right\}$,
where a word is called   reduced if it does not have subwords of the
form $x_i x_i^{-1}$ or $x_i^{-1} x_i$.

Any $m$-generated group $G$ is a factor group of $F_m$ by some
normal subgroup $N$. If there is a recursive (better yet, finite)
set of elements $\left\{r_{1}, \ldots, r_{k}, \ldots, \right\}$ of
$N$ that generate  $N$ as a  normal subgroup of $F_m$, then we use a
compact description of $G$ {\it by generators and defining
relators}:

$$G = \left\langle x_{1}, \ldots, x_{m} ~| ~r_{1}, \ldots, r_{k}, \ldots, \right\rangle$$

\noindent  and call $\left\{x_{1}, \ldots, x_{m}\right\}$ {\it
generators} of $G$, and $\left\{r_{1}, \ldots, r_{k}, \ldots,
\right\}$ {\it  defining relators} of $G$. Here ``generate  $N$ as a
normal subgroup of $F_m$" means that every element $u$ of $N$ can be
written as a (finite) product of conjugates of relators $r_{i}$:
~~$u = \prod_{k} h_k^{-1} r_{i_k} h_k$.

Given a presentation of a group $G$ as above, the {\it word problem}
for this presentation is:  given a word $w=w(x_{1}^{\pm 1}, \ldots,
x_{m}^{\pm 1})$ in the generators of $G$, find out whether or not $w
\in N$. (If $w \in N$, we say that $w =1$ in $G$.)

\section{An $(n,n)$-threshold Scheme}
\label{(n,n)}

Suppose we would like the dealer to distribute a $k$-column
$C=\left(
\begin{smallmatrix} c_{1}\\c_{2}\\ \cdot \\ \cdot \\ \cdot \\ c_{k}
\end{smallmatrix}\right)$  consisting of 0's and 1's among $n$
participants in such a way that the vector can be retrieved only
when all $n$ participants cooperate.  We begin by making a set of
group generators $X=\left\{x_{1}, \ldots, x_{m}\right\}$ public. The
scheme is as follows:

\begin{enumerate}

\item The dealer distributes over a secure channel to each participant $P_{j}$ a
set of words $R_{j}$ in the alphabet $X^{\pm 1}=\left\{x_{1}^{\pm
1}, \ldots, x_{m}^{\pm 1} \right\}$ such that each group
$G_{j}=\left\langle x_{1}, \ldots, x_{m} ~| ~R_{j}\right\rangle$ has
efficiently solvable word problem.

\item The dealer splits the secret bit column $C$ (the actual secret to be shared) into a sum
$C=\displaystyle \sum_{j=1}^{n}C_{j}$ of $n$ bit columns; these are
secret shares to be distributed.

\item The dealer then distributes words $w_{1j}, \ldots, w_{kj}$  in the
generators $x_{1}, \ldots, x_{m}$ over an open channel to each
participant $P_j$, $1\leq j \leq n$. The words are chosen so that
$w_{ij} \ne 1$ in  $G_{j}$  if $c_{i}=0$ and  $w_{ij}= 1$ in $G_{j}$
if $c_{i}=1$.

\item Participant $P_{j}$ then checks, for each $i$, whether  the word $w_{ij}=1$ in $G_{j}$ or not.
After that, each participant $P_{j}$ can make a column of 0's and
1's, $C_{j}= \left(
\begin{smallmatrix} c_{1j}\\c_{2j}\\ \cdot \\ \cdot \\ \cdot
\\ c_{kj} \end{smallmatrix}\right)$, by setting $c_{ij}=1$ if
$w_{ij}= 1$ in  $G_{j}$  and 0 otherwise.

\item  The participants then construct the secret by forming the column vector
$C=\displaystyle \sum_{j=1}^{n}C_{j}$, where the sum of the entries is taken modulo 2.

\end{enumerate}

In Step (5) of the above protocol,  the participants can use secure
computation  of a sum as proposed in \cite{GS} if they do not want
to reveal their individual column vectors, and therefore their
individual secret shares, to each  other. In order to implement the
protocol to compute a secure sum,  the participants should be able
to communicate over secure channels with one another. These secure
channels should be arranged in a circuit, say, $P_{1} \to P_{2} \to
\ldots \to P_{n} \to P_{1}$. Then the protocol to compute a secure
sum is as follows:

\begin{enumerate}

\item $P_{1}$ begins the process by choosing a random column vector $N_{1}$. He then sends to
$P_{2}$ the sum $N_{1}+C_{1}$.

\item  Each $P_{i}$, for $2\leq i \leq n-1$, does the following.  Upon receiving a column vector $C$
from participant $P_{i-1}$, each participant $P_{i}$ chooses a random column vector $N_{i}$ and adds
$N_{i}+C_{i}$ to $C$ and sends the result to $P_{i+1}$.

\item Participant $P_{n}$ chooses a random column vector $N_{n}$ and adds $N_{n}+C_{n}$ to the column
he has received from $P_{n-1}$ and sends the result to $P_{1}$. Now
$P_{1}$ has the column vector $\displaystyle
\sum_{i=1}^{n}(N_{i}+C_{i})$.

\item Participant $P_{1}$ subtracts $N_{1}$ from what he got from $P_n$; the
result now is the sum  $S = \sum_{1 \le i \le k} C_i ~+ ~\sum_{2 \le
i \le k} N_{i}$. (This step is needed for $P_{1}$ to preserve
privacy of his $N_{1}$, and therefore of his $C_{1}$, since $P_{2}$
knows $N_{1}+C_{1}$.) Then $P_{1}$ broadcasts $S$ to other
participants.

\item The participants then pool together to recover the secret. They do this
by each subtracting his random column vector $N_{i}$, $2 \le i \le
n$, from $S$.

\end{enumerate}

Thus, by using $n$ secure channels between the participants, the
participants are able to compute a secure sum in this secret sharing
scheme. For more on the computation of a secure sum see \cite{GS}.


\subsection{Efficiency.}  We  note that the dealer can efficiently  build a word
$w$ in the normal closure of $R_{i}$  as a product of arbitrary
conjugates of elements of $R_{i}$, so that $w= 1$ in  $G_{i}$.
Furthermore, if $G_{i}$ is a {\it small cancellation group} (see our
Section \ref{Platform}), then it is also easy to build a word $w$
such that $w \ne 1$ in $G_{i}$: it is sufficient to take care that
$w$ does not have more than half of any cyclic permutation of any
element of $R_{i}$ as a subword. See our Section \ref{Platform} for
more details. Finally, we note that in small cancellation groups
(these are the platform groups that we propose in this paper), the
word problem has a very efficient solution, namely, given a word $w$
in the generators of a small cancellation group $G$, one can
determine, in linear time in the length of $w$, whether or not $w=1$
in $G$.



\section{A $(t,n)$-threshold Scheme}
\label{(t,n)}

Here we propose a scheme that combines Shamir's idea  with our
scheme in Section \ref{(n,n)}.  As in Shamir's scheme, the secret is
an element $x\in \mathbb{Z}_{p}$, and the dealer chooses a
polynomial $f$ of degree $t-1$ such that $f(0)=x$. In addition the
dealer determines integers $y_{i}=f(i) ~(\mod p)$ that are to be
distributed to participants $P_{i}$, $1\leq i \leq n$. A set of
group generators $\left\{x_{1}, \ldots, x_{m}\right\}$ is made
public. We assume here that all integers $x$ and $y_{i}$ can be
written as $k$-bit columns. Then the scheme is as follows.

\begin{enumerate}

\item The dealer distributes over a secure channel to each participant $P_{j}$ a set of relators $R_{j}$
such that each group $G_{j}=\left\langle x_{1}, \ldots, x_{m} |
R_{j}\right\rangle$ has efficiently solvable word problem.

\item The dealer then distributes over open channels  $k$-columns\\
$b_{j}=\left( \begin{smallmatrix} b_{1j}\\b_{2j}\\ \cdot \\ \cdot \\
\cdot \\ b_{kj} \end{smallmatrix}\right)$, ~$1 \leq j \leq n$, of
words in $x_{1}, \ldots, x_{m}$ to each participant.  The words
$b_{ij}$ are chosen so that, after replacing them by bits (as usual,
``1" if $b_{ij}=1$ in the group $G_{j}$ and ``0" otherwise), the
resulting bit column represents the integer $y_{j}$.

\item Participant $P_j$ then checks, for each word $b_{ij}$, whether or not $b_{ij}=1$
in his/her group $G_{j}$, thus obtaining a binary representation of
the number $y_{j}$, and therefore recovering $y_{j}$.

\item  Each participant now has a point $f(i)=y_{i}$ of the polynomial. Using polynomial interpolation, any
$t$ participants can now recover the polynomial $f$, and hence the
secret $x=f(0)$.

\end{enumerate}

If $t \ge 3,$ then the last step of this protocol can be arranged in
such a way that participants do not have to reveal their individual
shares $y_{i}$ to each  other if they do not want to. Indeed, from
the Lagrange interpolation formula we see that

$$f(0) = \sum_{i=1}^t y_i \prod_{1 \le j \le t, ~j \ne i} \frac{-j}{i-j} .$$

Thus, $f(0)$ is a linear combination of private  $y_{i}$ with
publicly known coefficients $c_i = \prod_{1 \le j \le t, j \ne i}
\frac{-j}{i-j}$. If $t \ge 3,$ then this linear combination  can be
computed without revealing $y_{i}$, the same way the sum of private
numbers was computed in our Section \ref{(n,n)}.

In the special case $t = 2$, this yields an interesting problem.
Note that in the original Shamir's scheme, pairs  $(i, f(i))$ of
coordinates are sent to participants over secure channels, so that
the second coordinates are private, whereas the first coordinates
are essentially public because they just correspond to participants'
 numbers in an ordering that could be publicly known. This, however, does not have to be the case, i.e., the
first coordinates can be made private, too, so that the dealer sends
private points $(x_i, f(x_i))$ to participants. Then, for $t = 2$,
we have the following problem of independent interest:

\begin{problem}
Given that two participants, $P_1$ and $P_2$,  each  has a point
$(x_i, y_i)$ in the plane, is it possible for them to exchange
information in such a way that at the end, they both can recover an
equation of the line connecting their two points, but neither of
them can recover precise coordinates of the other participant's
point?
\end{problem}


\section{Why Use Groups?}
\label{Why?}

One might ask a natural question at this point: ``What is the
advantage of using groups in this scheme? Why not use just sets of
elements $R_{j}$ as long-term secrets, and then distribute elements
$w_{ij}$ that either match some elements of $R_{j}$ or not?" The
disadvantage of this procedure is that it will eventually compromise
the secrecy of $R_{j}$ because matching elements will have to be
repeated sooner or later. On the other hand, there are infinitely
many different words that are equal to 1 in a given group $G$. For
example, if $w=1$ in $G$, then also $\prod_{i} h_{i}^{-1}w h_{i}=1$
for any words $h_{i}$. Thus, the dealer can send as many words
$w_{ij}=1$ in $G$ to the participants as he likes,  without having
to repeat any word or update the relators $R_{j}$.

The question that still remains is whether some information about
relators $R_{j}$ may be leaked, even though the words distributed
over open channels will never match any words in $R_{j}$. This is an
interesting question of group theory; we address it, to some extent,
in our Section \ref{Tietze}.

%
%
%
%
%

\section{Platform Group}
\label{Platform}

In order for our scheme to be practical, we need each participant to
have a finite presentation of a group with efficiently solvable word
problem. Here we suggest small cancellation groups as a platform for
the protocol. For more information on small cancellation groups see
e.g. \cite{LS}.

  Let $F(X)$ be the free group on generators
$X=\left\{x_{1}, \ldots, x_{n}\right\}$.  A word $w(x_{1}, \ldots,
x_{n})=x_{i_{1}}^{\epsilon_{1}}\cdots x_{i_{n}}^{\epsilon_{n}}$,
where $\epsilon_{i}= \pm1$ for $1\leq i \leq n$, is called
\textit{cyclically reduced} if it is a reduced  word and
$x_{i_{1}}^{\epsilon_{1}}\neq x_{i_{n}}^{-\epsilon_{n}}$.


 A set $R$
containing cyclically reduced words is called \textit{symmetrized}
if it is closed under taking cyclic permutations and inverses. Given
a set $R$ of relators, a non-empty word $w\in F(X)$ is called a
\textit{piece} if there are two distinct relators $r_{1},r_{2}\in R$
such that $w$ is an initial segment of both $r_{1}$ and $r_{2}$;
that is, $r_{1}=wv_{1}$ and $r_{2}=wv_{2}$ for some $v_{1}, v_{2}\in
F(X)$ and there is no cancellation between $w$ and $v_{1}$ or $w$
and $v_{2}$.

In the definition below, $|w|$ denotes the lexicographic length of a
word $w$.

\begin{definition}Let $R$ be a symmetrized set of relators, and let $ 0< \lambda <1$.
A group $G=\left\langle X; R \right\rangle$ with the set $X$ of
generators and the set $R$ of relators is said to satisfy the
\text{small cancellation condition} $C^{'}(\lambda)$ if for every
$r\in R$ such that $r=wv$ and $w$ is a piece, one has $|w|< \lambda
|r|$. In this case, we say that $G$  belongs to the class
$C^{'}(\lambda)$.
\end{definition}

We propose groups that satisfy the small cancellation property
because groups in the  class $C^{'}(\frac{1}{6})$ have the  word
problem efficiently solvable by Dehn's algorithm. The algorithm is
straightforward:  given a word $w$, look for a subword of $w$ which
is a piece of a relator from $R$ of length  more than a half of the
length of the whole relator. If no such piece exists, then $w\neq 1$
in $G$. If there is such a piece, say $u$, then $r=uv$ for some
$r\in R$, where the length of $v$ is smaller than the length of $u$.
Replace the subword  $u$ by $v^{-1}$ in $w$, and the length of the
resulting word will become smaller than that of $w$. Thus, the
algorithm must terminate in at most $|w|$  steps. This (original)
Dehn's algorithm is therefore easily seen to have at most quadratic
time complexity with respect to the length of $w$. We note that
there is a slightly more elaborate version of Dehn's algorithm that
has linear time complexity.

We also note that a generic finitely presented group is a small
cancellation group (see e.g. \cite{Ol92}); this means, a randomly
selected set of relators will define a small cancellation group with
overwhelming probability. Therefore, to randomly select a small
cancellation group, the dealer in our  scheme can just take a few
random words of length $> 6$ and check whether the  corresponding
symmetrized set satisfies the condition for $C'(\frac{1}{6})$. If
not, then repeat.

\section{Tietze transformations: elementary isomorphisms}
\label{Tietze}

This section is somewhat more technical than the previous ones. Our
goal here is to show how to break long defining relators in a given
group presentation into short pieces by using simple
isomorphism-preserving transformations. This is useful because in a
small cancellation presentation (see our Section \ref{Platform})
defining relators tend to be long and, moreover, a word that is
equal to 1 in a presentation like that should contain a subword
which is a piece of a defining  relator  of length  more than a half
of the length of the whole relator. Therefore, exposing sufficiently
many words that are equal to 1 in a given presentation may leak
information about defining  relators. On the other hand, if all
defining  relators are short (of length 3, say), a word that is
equal to 1 in such a presentation is indistinguishable from random.

Long time ago, Tietze introduced isomorphism-preserving elementary
transformations that can be applied to groups presented by
generators and defining relators (see e.g. \cite{LS}). They are of
the following  types.

\begin{description}
\item[(T1)]
\emph{Introducing a~new generator}: Replace $\langle x_1,x_2,\ldots
\mid r_1, r_2,\dots\rangle$ by\\
 $\langle y, x_1,x_2,\ldots \mid
ys^{-1}, r_1, r_2,\dots\rangle$, where $s=s(x_1,x_2,\dots )$ is an
arbitrary word in the generators $x_1,x_2,\dots$.
\item[(T2)]
\emph{Canceling a~generator} (this is the converse of (T1)): If we
have a~presentation of the form $\langle y, x_1,x_2,\ldots \mid q,
r_1, r_2,\dots\rangle$, where $q$ is of the form $ys^{-1}$, and $s,
r_1, r_2,\dots $ are in the group generated by $x_1,x_2,\dots$,
replace this presentation by $\langle x_1,x_2,\ldots \mid r_1,
r_2,\dots\rangle$.
\item[(T3)]
\sloppy \emph{Applying an    automorphism}: Apply an automorphism of
the free group generated by $x_1,x_2,\dots $ to all the relators
$r_1, r_2,\dots$.
\item[(T4)]
\emph{Changing defining relators}: Replace the set $r_1, r_2,\dots$
of defining relators by another set $r_1', r_2',\dots$ with the same
normal closure. That means, each of  $r_1', r_2',\dots$ should
belong to the normal subgroup generated by $r_1, r_2,\dots$, and
vice versa.
\end{description}

Tietze has proved (see e.g. \cite{LS}) that two groups $\langle
x_1,x_2,\ldots \mid r_1, r_2,\dots\rangle$ and $\langle
x_1,x_2,\ldots \mid s_1, s_2,\dots\rangle$ are isomorphic if and
only if one can get from one of the presentations to the other by a
sequence of transformations \textup{(T1)--(T4)}.

For each Tietze transformation of the types \textup{(T1)--(T3)}, it
is easy to obtain an explicit isomorphism (as a mapping on
generators) and its inverse. For a Tietze transformation of the type
\textup{(T4)}, the isomorphism is just the identity map. We would
like here to make Tietze transformations of the type \textup{(T4)}
recursive, because {\it a priori} it is not clear how one can
actually apply these transformations. Thus, we are
 going to use the following recursive version of \textup{(T4)}:

\medskip

\noindent {\bf (T4$'$)} In the set $r_1, r_2,\dots$, replace some
$r_i$ by one of the:  $r_i^{-1}$,  $r_i r_j$, $r_i r_j^{-1}$, $r_j
r_i$, $r_j r_i^{-1}$, $x_k^{-1} r_i x_k$, $x_k r_i x_k^{-1}$, where
$j \ne i$, and $k$ is arbitrary.
\medskip

Now we explain how the dealer can break down given defining relators
into short pieces. More specifically, he can replace a given
presentation by an isomorphic presentation where all defining
relators have length at most 3.  This is easily achieved by applying
transformations \textup{(T1)} and  \textup{(T4$'$)}, as follows. Let
$\Gamma$ be a presentation $\langle x_1,\dots , x_k;
r_1,...,r_m\rangle$. We are going to obtain a different, isomorphic,
presentation by using Tietze transformations of types \textup{(T1)}.
 Specifically, let, say,  $r_1=x_i x_j u$, $1 \le i,j \le k$.
We introduce a new generator $x_{k+1}$  and a new relator
$r_{m+1}=x_{k+1}^{-1}x_i x_j$. The presentation \\
$\langle x_1,\dots ,x_k, x_{k+1}; r_1,\dots ,r_m, r_{m+1}\rangle$ is
obviously isomorphic to $\Gamma$. Now if we replace $r_1$ with
$r_1'=x_{k+1} u$, then the presentation $\langle x_1,\dots ,x_k,
x_{k+1}; r_1',\dots ,r_m, r_{m+1}\rangle$ will again be isomorphic
to $\Gamma$, but now the length of one of the defining relators
($r_1$) has decreased by 1. Continuing in this manner, one can
eventually obtain a presentation where all relators have length at
most 3, at the expense of introducing more generators.

 We conclude this section with a simple  example, just to illustrate
how Tietze transformations can be used to cut relators into pieces.
In this example, we start with a presentation having two relators of
length 5 in 3 generators, and end up with a presentation having 5
relators of length 3 in 6 generators. The symbol $\cong$ below means
``is isomorphic to".

\medskip

\noindent {\bf Example.}  $\langle x_1, x_2, x_3 ~\mid ~x_1^2x_2^3,
~x_1x_2^2x_1^{-1}x_3 \rangle ~\cong ~\langle x_1, x_2, x_3, x_4
~\mid ~x_4=x_1^2, ~x_4x_2^3, ~x_1x_2^2x_1^{-1}x_3 \rangle\\
\cong \langle x_1, x_2, x_3, x_4, x_5 ~\mid  ~x_5=x_1x_2^2,
~x_4=x_1^2, ~x_4x_2^3, ~x_5x_1^{-1}x_3 \rangle\\
\cong \langle x_1, x_2, x_3, x_4, x_5, x_6 ~\mid  ~x_5=x_1x_2^2,
~x_4=x_1^2, ~x_6=x_4x_2, ~x_6x_2^2, ~x_5x_1^{-1}x_3 \rangle$.

\medskip

We note that this procedure of breaking relators into pieces of
length 3 increases the total length of relators by at most the
factor of 2.

\section{Conclusions}

We have proposed a two-stage $(t,n)$-threshold secret sharing scheme
where long-term secrets are distributed to participants over secure
channels, and then shares of the actual secret can be distributed
over open channels.

Our scheme  has the same distributed secret as Shamir's scheme does,
but rather than sending shares of a secret over secure channels, we
send the integers in disguise (as tuples of words in a public
alphabet) over open channels, and then participants use
group-theoretic methods to recover the integers. This scheme has the
following useful advantages over Shamir's original scheme:

\begin{itemize}
\item The actual secret need not be  distributed over
secure channels and, furthermore, once the long-term private
information is distributed to all participants (over secure
channels), several different secrets can be distributed without
updating the  long-term private information.

\item While recovering the secret,  participants do not have to reveal their shares
to each  other if they do not want to.
\end{itemize}

\bibliographystyle{plain}
\bibliography{XBib}
\end{document}